\documentclass[10pt,twoside,a4paper, twocolumn]{article}
\newcommand{\Title}{Self-assembling kinetics:\\ Accessing a new design space via differentiable statistical-physics models} 
\newcommand{\Author}{Goodrich and King et al} 

\pdftrailerid{} 
\usepackage[affil-sl]{authblk}
\renewcommand{\thefootnote}{\fnsymbol{footnote}}

\usepackage{titlesec}
\titleformat*{\section}{\large \sc \bfseries}
\titleformat*{\subsection}{}
\titleformat*{\subsubsection}{\sl}

\usepackage{xcolor}
\definecolor{maroon}{rgb}{0.5, 0.0, 0.0}
\definecolor{PUSblue}{HTML}{005FA8}
\usepackage{hyperref} \hypersetup{colorlinks = true, citecolor = PUSblue}

\usepackage{lipsum}
\usepackage{fancyhdr}
\usepackage{amssymb}
\usepackage{graphicx}
\usepackage{hyperref}
\usepackage{multirow}

\usepackage[utf8]{inputenc} 
\usepackage[]{kpfonts}
\usepackage[T1]{fontenc}

\usepackage{graphicx}
\usepackage{booktabs}
\usepackage{longtable}
\usepackage{siunitx}
\usepackage[nottoc]{tocbibind} 
\usepackage[french,english]{babel}
\usepackage{amsmath}

\usepackage{draftwatermark}
\SetWatermarkText{\sc DRAFT}
\SetWatermarkScale{1}
\SetWatermarkColor[gray]{0.85}

\usepackage{lineno}

\usepackage[tmargin=1in,bmargin=1in,lmargin=0.9 in,rmargin=0.9in,headheight=42pt]{geometry} 

\usepackage{abstract}

\fancypagestyle{firststyle}
{
   \fancyhf{}
}

\newcommand\blfootnote[1]{%
  \begingroup
  \renewcommand\thefootnote{}\footnote{#1}%
  \addtocounter{footnote}{-1}%
  \endgroup
}

\renewcommand{\v}[1]{\ensuremath{\mathbf{#1}}} 
 
\newcommand{\guv}[1]{\ensuremath{\mbox{\boldmath$ \hat{#1} $}}} 

\title{Self-assembling kinetics: Accessing a new design space via differentiable statistical-physics models}

\author{%
Carl P. Goodrich$^{a, b }$\thanks{Corresponding author: {\texttt{carl.goodrich@ist.ac.at}}}, 
 Ella M. King$^c$\thanks{Co-first author}, Samuel S. Schoenholz$^{d}$, Ekin D. Cubuk$^{d}$, and Michael Brenner$^{a, c, d}$%
} 
\usepackage{lmodern}
\begin{document}
\date{}
\twocolumn[
  \begin{@twocolumnfalse}
    \maketitle
\thispagestyle{firststyle}

\selectlanguage{english}
\begin{abstract}
The inverse problem of designing component interactions to target emergent structure is fundamental to numerous applications in biotechnology, materials science, and statistical physics~\cite{whitesides1991molecular, douglas1998host, lohmeijer2002supramolecular, hartgerink1996self}. Equally important is the inverse problem of designing emergent  kinetics, but this has received considerably less attention. 
Using recent advances in automatic differentiation, ~\cite{ad,jax,jaxmd2019,Schoenholz:2019tq}
we show how kinetic pathways can be precisely designed by directly differentiating through statistical-physics models, namely free energy calculations and molecular dynamics simulations. We consider two systems that are crucial to our understanding of structural self-assembly: bulk crystallization and small nanoclusters. In each case we are able to assemble precise dynamical features. Using gradient information, we manipulate interactions among constituent particles to tune the rate at which these systems yield specific structures of interest. Moreover, we use this approach to learn non-trivial features about the high-dimensional design space, allowing us to accurately predict when multiple kinetic features can be simultaneously and independently controlled. These results provide a concrete and generalizable foundation for studying non-structural self-assembly, including kinetic properties as well as other complex emergent properties, in a vast array of systems.
\end{abstract}

  \end{@twocolumnfalse}
]
\saythanks
\blfootnote{$^a$ School of Engineering and Applied Sciences, Harvard University, Cambridge, MA 02138, USA}
\blfootnote{$^b$ Institute of Science and Technology Austria. A-3400 Klosterneuburg, Austria}
\blfootnote{$^c$ Physics Department, Harvard University, Cambridge, MA 02138, USA}
\blfootnote{$^d$ Google Research, 1600 Amphitheatre Pkwy, Mountain View, CA}

\section{Introduction}

Kinetic features are a critical component of a wide range of biological and material functions. The complexity seen in biology could not be achieved without control over intricate dynamic features. Viral capsids are rarely infectious if they assemble too quickly~\cite{virusassembly}, protein folding requires deft navigation and control of kinetic traps~\cite{protein}, and crystal growth is largely controlled by the relative rates of different nucleation events~\cite{kincrys}. However, the design space of kinetic features remains largely unexplored. Self-assembly traditionally focuses on tuning structural properties, and while there have been significant successes in developing complex structural features~\cite{dnaorigami, whitesides1991molecular, douglas1998host, stupp1997supramolecular, lohmeijer2002supramolecular, hartgerink1996self}, there has been little exploration of dynamical features in the same realm. We demonstrate quantitative control over assembly rates and transition rates in canonical soft matter systems. In doing so, we begin to explore the vast design space of dynamical materials properties for the first time.

To intelligently explore the design space of dynamical features, we rely on recent advances in Automatic Differentiation (AD)~\cite{Bryson:1962is,bryson1975applied,Linnainmaa:1976vp,rumelhart1986learning,ad}, a technique for efficiently computing exact derivatives of complicated functions automatically, along with advances in efficient and powerful implementations of AD algorithms~\cite{NEURIPS2019_9015,tensorflow2015-whitepaper,jax} on sophisticated hardware such as GPUs and TPUs. While the development of these techniques is driven by the machine learning community, a distinct, non-data-driven approach is emerging that combines AD with traditional scientific simulations. Rather than training a model using preexisting data, AD connects observables to physical parameters by directly accessing physical models. This enables intelligent navigation of high-dimensional phase spaces. Initial examples utilizing this approach include optimizing the dispersion of photonic crystal waveguides~\cite{Minkov:2020vc}, the invention of new coarse grained algorithms for solving nonlinear partial differential equations~\cite{bar2019learning}, the discovery of molecules for drug development~\cite{mccloskey2020machine} and greatly improved predictions of protein structure~\cite{senior2020improved}.

Here, we begin to explore an entirely new class of complex inverse design problems. Using AD to train well-established, statistical-physics based models, we design materials for \textit{dynamic}, rather than structural, features. We also use AD to gain theoretical insights into this design space, allowing us to predict the extent of designability in different systems. Automatic differentiation is an essential component of this approach because we rely on gradients to connect physical parameters to complex emergent behavior. While there are other approaches for obtaining gradient information, e.g. finite difference approximations, AD calculates exact derivatives and, more importantly, can efficiently handle large numbers of parameters. Furthermore, the theoretical insights we develop rely on accurate calculations of the Hessian matrix of second derivatives, for which finite-difference approaches are insufficient.

We start in Sec.~\ref{sec:honeycomb_crystallization} by considering the bulk crystallization of identical particles into both honeycomb and triangular lattices. By differentiating over entire molecular dynamics trajectories with respect to interaction parameters, we are able to tune their relative crystallization rates. We proceed with a more experimentally relevant example in Sec.~\ref{sec:colloidal_clusters}, where we consider small nanoclusters of 7 colloidal particles that can transition between structural states. By optimizing over the transition barriers, we obtain precise control over the transition dynamics. Together, Secs.~\ref{sec:honeycomb_crystallization} and \ref{sec:colloidal_clusters} demonstrate that kinetic features can be designed in classic self-assembly systems while maintaining the constraint of specific structural motifs. The use of automatic differentiation provides a way to directly access the design space of kinetic features by connecting emergent dynamics with model parameters. This connection reveals insight into the high-dimensional design landscape, enabling us to make accurate predictions in Sec.~\ref{sec:colloidal_clusters}\ref{sec:simultaneous_transitions} regarding when large numbers of kinetic features of the colloidal clusters can be controlled simultaneously. Finally, we conclude in Sec.~\ref{sec:discussion} with a discussion of both the specific and broad impact of these results. 

\section{Tuning assembly rates of honeycomb crystals \label{sec:honeycomb_crystallization}} 
We start with the classic self-assembly problem introduced by Rechtsman et al.~\cite{torquato,torquato_hon} of assembling a honeycomb lattice out of identical isotropic particles. The honeycomb lattice is a two-dimensional analog of the diamond lattice, which exhibits desirable mechanical and photonic properties, but its low density makes assembling such a crystal a particularly difficult problem. Nevertheless, Rechtsman et al.~\cite{torquato} successfully designed a honeycomb lattice using a potential of the form 
\begin{equation} \label{eq:honpot}
U(r) = \frac{b_0}{r^{12}} -  \frac{a_0}{r^{10}} + a_1 e^{-a_2 r} - b_1 e^{-b_2(r - a_3)^2}.
\end{equation}
They fixed the parameters $b_0 = 5$, $b_1 = 0.4$, and $b_2 = 40$, and optimized $a_0$, $a_1$, $a_2$, and $a_3$ using simulated annealing~\cite{torquato,torquato_hon} at zero temperature. 
We optimize the same set of four free parameters. However, we optimize via automatic differentiation rather than simulated annealing, and work towards a different goal: tuning the \emph{rate} of assembly at finite temperature, while still enforcing the structural constraint that a honeycomb lattice is assembled. 

\begin{figure}[h!tpb]
	\centering
	\includegraphics[width=\linewidth]{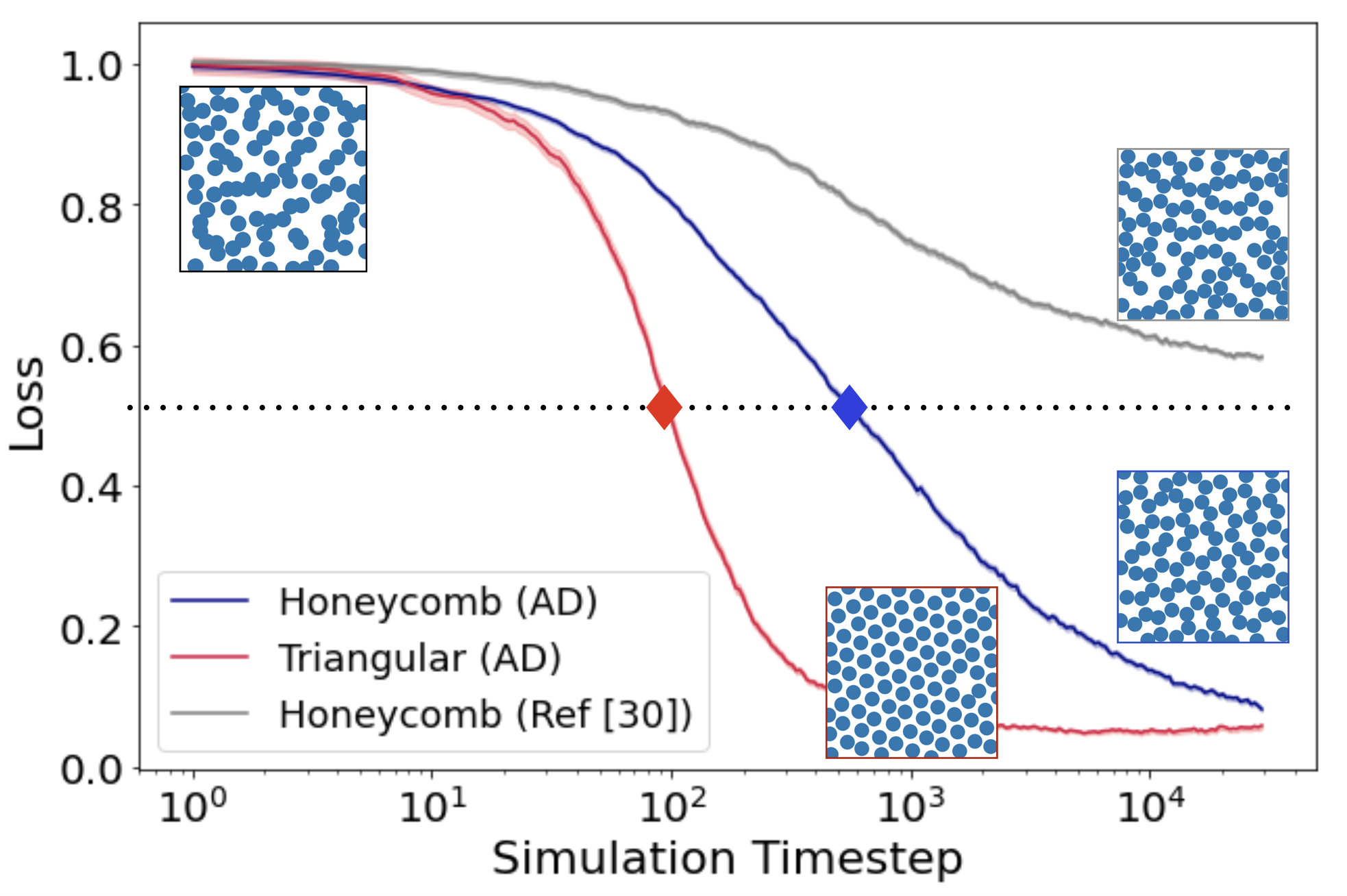}
	\caption{\label{fig:assembly_proc} Assembly process for honeycomb and triangular lattices. Two different loss functions are pictured: a honeycomb loss and a triangular loss. The honeycomb loss is shown in blue for the parameters found using automatic differentiation and in gray for the parameters found by Rechtsman et al~\cite{torquato_hon}, whereas the triangular loss is shown in red. Each curve is the result of 200 independent simulations, while the triangular lattice simulations are performed under a higher density than the honeycomb lattice simulations. The images on the right-hand side are representative examples of the assembled lattice structures. The bottom two images use the same set of parameters as found using automatic differentiation but have different volume fractions. The image on the upper left shows a sample random initial configuration.  In order to extract assembly rates, we note the time at which the lattices are half-assembled, demarcated with a diamond. }
\end{figure}

We consider a system of $N=100$ particles that interact via Eq.~\eqref{eq:honpot} and evolve under overdamped Langevin dynamics at constant temperature (Appendix~\ref{sec:Methods}). 
The blue curve in Fig.~\ref{fig:assembly_proc} shows the assembly dynamics for a specific set of parameters. Specifically, it shows a ``honeycomb loss function", $L_\mathrm{H}$, that measures the onset of crystallization by identifying the most ``honeycomb-like" region, which is a leading indicator of bulk crystallization (Appendix~\ref{sec:methods_crystallization}). $L_\mathrm{H}\approx 1$ for a randomly arranged system and decreases to $L_\mathrm{H}= 0$ for a perfect honeycomb lattice. Going forward, we will define the crystallization rate $k_\mathrm{H} \equiv 1/t_\mathrm{H}$, where $t_\mathrm{H}$ is the time at which $L_\mathrm{H}$ drops below 0.5 (indicated by the blue diamond in Fig.~\ref{fig:assembly_proc}). 

The parameters in Fig.~\ref{fig:assembly_proc} were not chosen randomly but are the results of an initial optimization procedure. As explained in Appendix~\ref{sec:methods_crystallization},
we use AD to calculate the gradient of $L_\mathrm{H}$ after 3,300 simulation timesteps using randomly chosen initial parameters. With the gradient in hand, we iteratively minimize $L_\mathrm{H}$ using a standard optimization procedure.  
Though not targeting a dynamical rate, these results are the first example of AD-based optimization over a molecular dynamics simulation. Note that this is slightly different than the aim of Rechtsman et al.~\cite{torquato,torquato_hon}, who were focused on the zero-temperature ground state, and not surprisingly their results (grey curve in Fig.~\ref{fig:assembly_proc}) are not optimal for finite-time, finite-temperature assembly.

While the system forms a honeycomb lattice at low volume fractions, the same system with the same interaction parameters will assemble into a triangular lattice at a higher volume fractions. The ``triangular loss function", $L_\mathrm{T}$, shown in red in Fig.~\ref{fig:assembly_proc}, can similarly be used to define a second crystallization rate $k_\mathrm{T}$.
These rates are measured relative to intrinsic time scales that are nontrivially coupled to the potential. To regularize the intrinsic timescale, we seek to control $k_\mathrm{H}$ {\it relative} to $k_\mathrm{T}$.

We use automatic differentiation over simulations of the assembly process to optimize for our desired assembly dynamics. To understand how we perform this optimization, consider the function

\begin{equation}
\mathcal{S}_\mathrm{MD} \left(a, R, \rho, t \right)
\end{equation}
which runs a Molecular Dynamics simulation for $t$ simulation steps at density $\rho$, starting from an initial set of positions $R$, and with interactions given by ~\eqref{eq:honpot} with parameters $a = \{ a_0,a_1,a_2,a_3\}$. The function returns the final set of positions $R_t$ after $t$ simulation timesteps. 
Using this notation, Fig.~\ref{fig:assembly_proc} shows $L_\mathrm{H}(t) \equiv \mathcal{L}_\mathrm{H}\left(\mathcal{S}_\mathrm{MD} \left(a, R_0, \frac{2}{3} \rho_\mathrm{T}, t \right) \right)$ in blue and  $L_\mathrm{T}(t) \equiv \mathcal{L}_\mathrm{T}\left(\mathcal{S}_\mathrm{MD} \left(a, R_0, \rho_\mathrm{T}, t \right) \right)$ in red, where $\mathcal{L}_\mathrm{H,T}(R)$ are functions that return the respective loss function for a set of positions over time.

The calculation of $L_\mathrm{H}(t)$ (or $L_\mathrm{T}(t)$) is analogous to a simple feed-forward neural network with $t$ hidden layers corresponding to each time step and where $\mathcal{L}_\mathrm{H}$ is applied to the last hidden layer to obtain the output. However, rather than using a set of variable weights and biases to move from one layer to the next, we use the discretized equations of motion to integrate the dynamics. Furthermore, the variables that we train over, $a$, are physical parameters (e.g. that define the interaction potential) rather than the millions of weights and biases that typically comprise a neural network. By using AD to propagate the gradient of, for example, $L_\mathrm{H}(t)$ through each simulation step to obtain $\frac{d}{da}L_\mathrm{H}(t)$, we are able to {\it train} a stochastic Molecular Dynamics simulation in a way similar to standard neural networks. Importantly, however, this is a data-free approach with ``baked-in physics" where results are not only interpretable but correspond to physical parameters.

\begin{figure}[h!tpb]
	\centering
	\includegraphics[width=\linewidth]{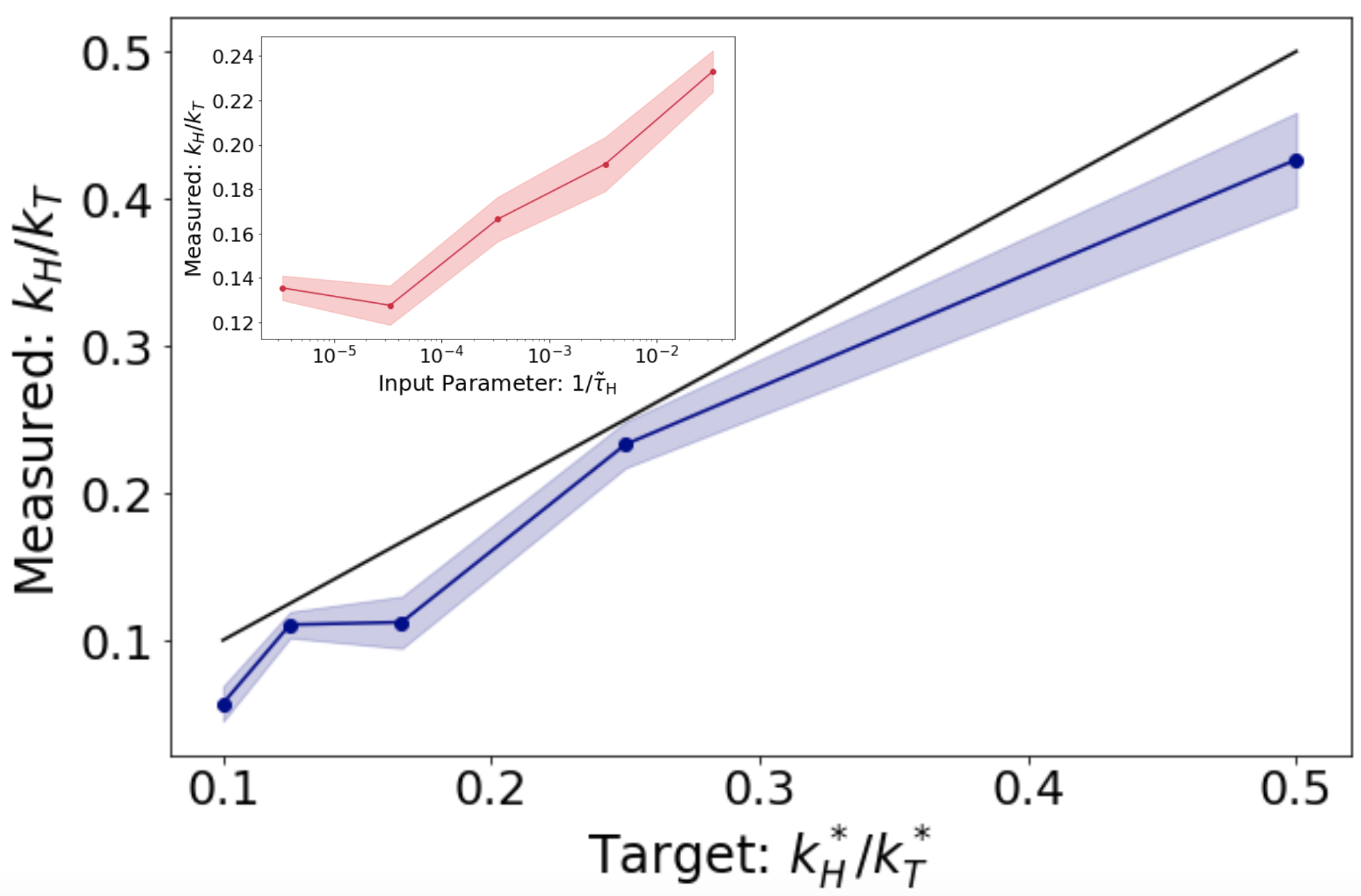}
	\caption{\label{fig:kHkT_res} 
	Ratio of assembly rates as a function of target ratio. Using forward-mode AD, we obtain non-trivial control over assembly rates (black line indicates perfect agreement). The inset shows the results using backward-mode AD (Appendix~\ref{sec:methods_crystallization}). 
	The backwards mode results are achieved by adjusting a tuning parameter $1/\tilde{\tau}_H$. By changing $1/\tilde{\tau}_H$, we obtain indirect control over the assembly rates by differentiating over only the final 300 simulation steps. Though less precise, this approach is more scalable to systems with many parameters.
	}
	
\end{figure}


We now seek to find parameters, $a$, such that the rate of the honeycomb lattice at density $\rho_H$ is $k_H^*$ and the rate of the hexagonal lattice at density $\rho_T$ is $k_T^*$. This is equivalent to minimizing the squared-loss, $L = (L_T(t^*_T) - \frac1 2)^2 + (L_H(t_H^*) - \frac1 2)^2$. The gradient $\frac{d L}{d a}$, which involves differentiating over two ensembles of molecular dynamics simulations, is used in conjunction with the RMSProp~\cite{rmsprop} stochastic optimization algorithm to minimize $L$. Figure~\ref{fig:kHkT_res} shows the results of this procedure, where we have fixed $k_\mathrm{T}^*=0.01$ and varied $k_\mathrm{H}^*$. While the ratio $k_\mathrm{H}/k_\mathrm{T}$ is not identical to the target ratio, we have obtained nontrivial control over the rates using only 4 parameters. 

These results were generated using ``forward mode" AD, in which memory usage is independent of the length of the simulation but computation time scales with the number of parameters. Forward-mode AD quickly becomes extremely time-intensive for more complex systems. A more scalable approach is to use ``reverse mode" AD, where the computation time scales favorably with the number of parameters. In the reverse mode formulation, however, the entire simulation trajectory must be stored in memory, which severely limits the size and length of the simulation. While techniques such as gradient rematerialization~\cite{remat} can be used to mitigate this, we have developed an alternative strategy (Supporting Information)
where only the last 300 time steps are differentiated over (inset to Fig.~\ref{fig:kHkT_res}). This method provides a more robust solution for large scale optimization and is similar in spirit to truncated backpropagation used in language modeling~\cite{langmod} and metalearning~\cite{metalearn}.

\section{Tuning transition rates in colloidal clusters \label{sec:colloidal_clusters}}

\begin{figure*}[h!tpb]
	\centering
	\includegraphics[width=\linewidth]{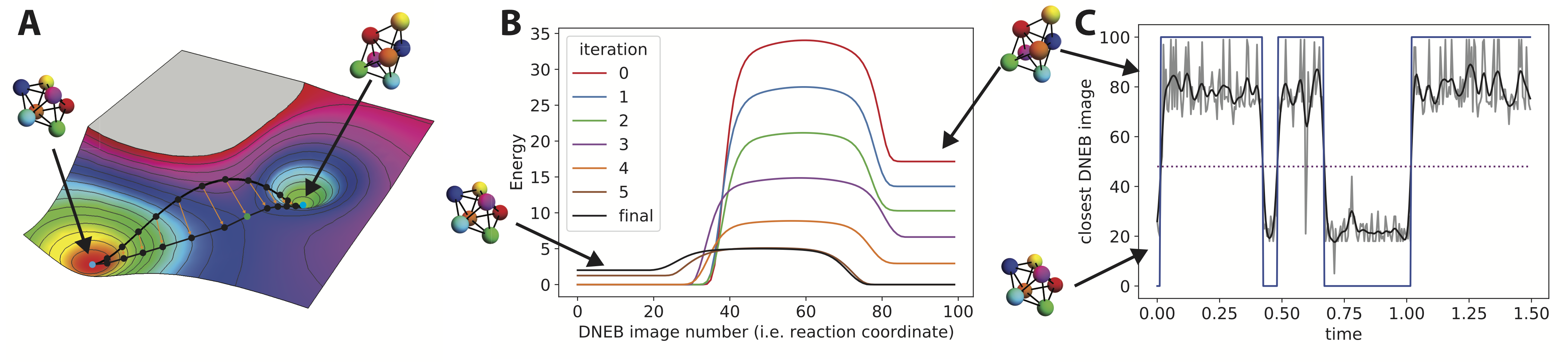}
	\caption{\label{fig:dneb_image} 
	a) Illustration of the Doubly Nudged Elastic (DNEB) method~\cite{Trygubenko:2004ip} for finding the transition state between two local energy minima, which in our case correspond to meta-stable 7-particle clusters. Given such an energy landscape, we construct a series of $n_I=100$ images that span the two minima and are connected by high-dimensional springs. Keeping the two endpoints fixed to the minima, the energy of the ensemble is minimized collectively to obtain the steepest descent path over the saddle point, or transition state. See Appendix~\ref{sec:methods_transitions} for more details. 
	B) Energy (in $k_BT$) along the steepest descent path after successive iterations of the optimization algorithm. For clarity, the minimum energy along the path is subtracted. Note that the regions of constant energy at the beginning and end of the path correspond to global rotations of the cluster.
	C) Verification of the transition rates via Molecular Dynamics (MD) simulation. Periodic snapshots from MD simulations are mapped to the steepest descent path found through the DNEB calculation (grey). Noise is reduced using a Butterworth low-pass filter (black), and then binarized (blue). The dashed line represents the threshold for binarizing the signal and corresponds to the image with the highest energy, i.e. $R_t$. Transition rates are calculated from the dwell times of the binarized signal.
	} 
\end{figure*}


We now turn to another ubiquitous dynamical feature in physical and biological systems. Spontaneous transitions between distinct structural configurations are crucially important in many natural processes from protein folding to allosteric regulation and trans-membrane transport, and the rates of these transitions are critical for their respective function. 
While the prediction of transition rates from energy landscapes is well-studied~\cite{Kramers:1940jm,Hanggi:1990en}, there have been few attempts to {\it control} transition rates. This is largely because a general understanding of how changes to interactions affect emergent rates is lacking.
Such an understanding, as well as the inverse problem of computing energy landscapes for a specific rate, are essential both for understanding biophysical functionality and for exploring feature space in physical systems.



In this section, we focus on a simple system that exhibits such transitions and can be realized \textit{experimentally}.
Clusters of micron scale colloidal particles are ideal model materials both theoretically and in the lab. Furthermore, advances in DNA nanotechnology~\cite{Mirkin:1996em,Valignat:2005ed,Biancaniello:2005ie} have enabled precise control over binding energies between specified particles by coating the colloids with specific DNA strands. 
As demonstrated by Hormoz and Brenner~\cite{Hormoz:2011ir} and Zeravcic et al.~\cite{Zeravcic:2014it}, and confirmed experimentally by Collins~\cite{CollinsThesis2014}, such control over the binding energies allows for high-yield assembly of specific clusters under thermal noise. The assembly of one stable cluster over the others is obtained by choosing the binding energies $B_{\alpha \beta}$ between particle pairs $(\alpha,\beta)$ so that the target structure is the ground state. While there is significant interest in avoiding kinetic traps in order to maximize yield~\cite{Dunn:2015ih}, there have been few attempts to control persistent kinetic features of these systems.

We aim to control the rate of transitioning between distinct states. We do this in two ways: first by simultaneously tuning the energy of the connecting saddle point {\it and} the state energies themselves, and second by directly tuning the Kramers approximation for the transition rates, which also takes the curvature of the energy landscape into account. We then consider the question of how many different kinetic features can be tuned simultaneously. To address this, we develop a constraint-based theory analogous to rigidity percolation that predicts when simultaneous control is possible and reveals insight into the nature of the design landscape.

For concreteness going forward, we consider clusters of $N=7$ spheres that interact via a short-ranged Morse potential with binding energy $B_{\alpha\beta}$ and with dynamics given by the overdamped Langevin equation with friction coefficient $\gamma$ (Appendix~\ref{sec:Methods}).
For positive $B_{\alpha\beta}$, Arkus et al.~\cite{Arkus:2011gda} identified all possible stable states up to permutations, each with 15 stabilizing contacts and no internal floppy modes. From these, we pick states that are separated by only a single energy barrier and use the $\frac{1}{2}N(N-1) = 21$ different $B_{\alpha\beta}$ as adjustable parameters to control the transition kinetics.

The transition rate from state $i$ to state $j$ is
\begin{equation}\label{eq:TST}
k_{ij} = \nu_{ij} e^{-\beta \Delta E_{it}}
\end{equation}
where $\Delta E_{it}$ is the energy barrier, and $\beta\equiv (k_BT)^{-1}$ with $k_B$ the Boltzmann constant and $T$ the temperature. $\nu_{ij}$ is a non-trivial prefactor that we can approximate using Kramers theory~\cite{Kramers:1940jm,Hanggi:1990en}:
\begin{equation} \label{eq:Kramers}
\nu_{ij} = \frac{{ \omega_b}}{2\pi\gamma}\frac{\prod_\ell {\omega_\ell^A}}{\prod_{\ell^\prime} {\omega_{\ell^\prime}^S}}
\end{equation}
where $\omega_b$ and $\{\omega_{\ell^\prime}^S\}$ are the frequencies of the unstable and stable vibrational modes, respectively, at the saddle point, and $\{\omega_\ell^A\}$ are the frequencies at state $i$.
We will demonstrate control over the transition kinetics in two ways: 1) by controlling the forwards and backwards energy barriers $\Delta E_{it}$ and $\Delta E_{jt}$, and 2) by controlling directly the forwards and backwards transition rates $k_{ij}$ and $k_{ij}$ using the Kramers approximation.


In order to proceed, we first have to find the transition state separating states $i$ and $j$. 
This is done using the Doubly Nudged Elastic Band (DNEB) method~\cite{Trygubenko:2004ip}, which is illustrated in Fig.~\ref{fig:dneb_image}A and described in Appendix~\ref{sec:methods_transitions}. The transition state, along with the initial and final states, allows us to calculate the energy barriers as well as the vibrational frequencies necessary for ~\eqref{eq:Kramers}.  

With this in hand, we next compute a loss function, $L$. When targeting the energy barriers, we use
\begin{align} \label{eq:barrierLoss}
L = \left(\Delta E_{it} - \Delta E_{it}^*\right)^2 + \left(\Delta E_{jt} - \Delta E_{jt}^*\right)^2
\end{align}
where $\Delta E_{it}^*$ and $\Delta E_{jt}^*$ are the desired energy barriers. When targeting the transition rates directly, we use
\begin{align} \label{eq:rateLoss}
L = \left(k_{ij} - k_{ij}^*\right)^2 + \left(k_{ji} - k_{ji}^*\right)^2
\end{align}

where $k_{ij}$ and $k_{ij}$ are the Kramers rates and $k_{ij}^*$ and $k_{ij}^*$ are the targets. Finally, we optimize $L$ using reverse-mode automatic differentiation to calculate the 21 derivatives $\left.\frac{d L}{d B_{\alpha\beta}}\right|_{R_i,R_j,R_t}$, where $R_{i,j,t}$ are the positions of the particles in the respective states. This gradient is fed into a standard optimization algorithm (RMSProp~\cite{rmsprop}) to minimize the loss, and $R_t$ is recalculated only as necessary (in practice every 50 optimization steps, see Appendix~\ref{sec:methods_transitions}, which we collectively refer to as a single ``iteration"). The entire process (18 iterations) takes roughly three minutes on a Tesla P100 GPU.

\begin{figure}[h!tpb]
	\centering
	\includegraphics[width=\linewidth]{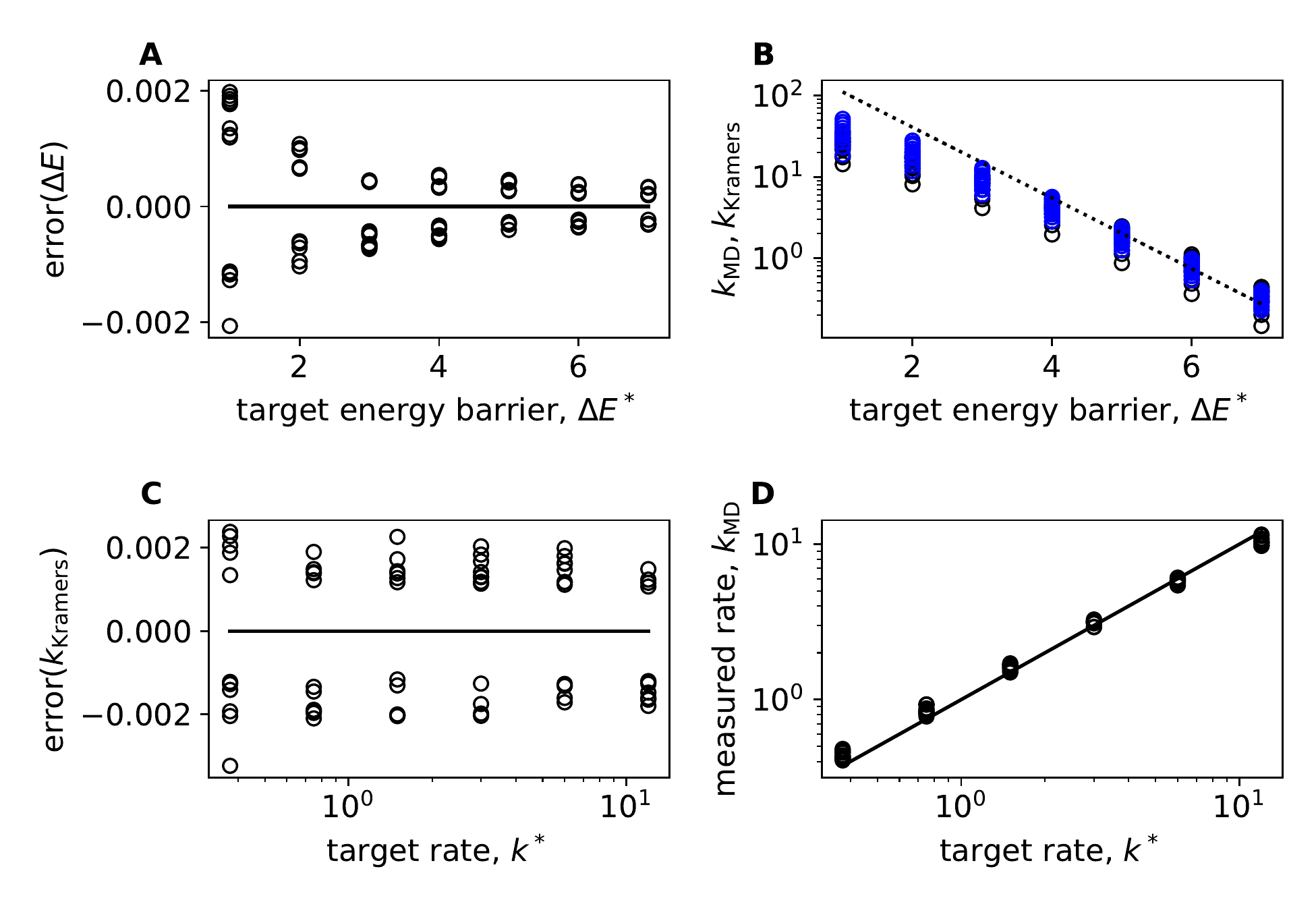}
	\caption{\label{fig:dneb_results}Training energy barriers and transition rates. 
	A) After 18 iterations of our procedure, the energy barrier $\Delta E$ is trained to within $0.2\%$ the target barrier $\Delta E^*$. Shown is the error ($\mathrm{error}(\Delta E) \equiv (\Delta E - \Delta E^*)/\Delta E^*$) as a function of the target. 
	B) The transition rates measured via MD simulation ($k_\mathrm{MD}$, black data) are not exactly proportional to $e^{-\beta \Delta E^*}$ (the dashed line has a slope of $-\beta$), indicating that the prefactor $\nu$ in Eq.~\eqref{eq:TST} is not constant. However, this prefactor is captured reasonably well by the Kramers approximation ($k_\mathrm{Kramers}$, blue data).
	C) After 18 iterations of training on the Kramers rate $k_\mathrm{Kramers}$, we obtain an accuracy within $0.2\%$ of the target rate $k^*$. 
	D) The rates measured via MD simulation $k_\mathrm{MD}$ agree very well with the target rate $k^*$ (the solid line corresponds to $k_\mathrm{MD}=k^*$). Thus, we are able to accurately and quantitatively design the transition kinetics of the 7-particle clusters we consider.
	}
\end{figure}

As a first example, we use Eq.~\eqref{eq:barrierLoss} to target energy barriers of $3k_BT$ and $5 k_BT$ between two of the stable $N=7$ particle clusters. Figure~\ref{fig:dneb_image}B shows the energy along the reaction coordinate, i.e. the steepest descent path that connects the two meta-stable states via the transition state, after various iterations of the above procedure. The final energy differences of $\Delta E_{it} = 2.998 k_BT$ and $\Delta E_{jt} = 5.001 k_BT$ are both within $0.1\%$ of their target, and can be further refined by continuing the optimization procedure. Figure~\ref{fig:dneb_results}A shows the results of 49 distinct target combinations and demonstrates that similar accuracy can be obtained over a wide range of energies. 

The connection between the energy barriers and the resulting transition rates are predicted using the Kramers approximation, which is validated via Molecular Dynamics (MD) simulations where we extract the rates directly (Fig.~\ref{fig:dneb_image}C and Appendix~\ref{sec:methods_transitions}). 
Figure~\ref{fig:dneb_results}B shows both rates ($k_\mathrm{Kramers}$ and $k_\mathrm{MD}$) as a function of the target energy barrier. While the validation agrees well with the Kramers prediction, the prefactor $\nu$ is not constant, underscoring the difficulty of targeting energy barriers as a proxy for transition rates without a quantitative model for $\nu$. 

An alternative is to target $k_\mathrm{Kramers}$ directly, which we do using the rate-based loss function (Eq.~\eqref{eq:rateLoss}). As shown in Fig.~\ref{fig:dneb_results}C, we obtain a typical error of $0.2\%$ after the optimization. As before, $k_\mathrm{MD} \approx k_\mathrm{Kramers}$, and Fig.~\ref{fig:dneb_results}D shows that the full dynamics of the MD simulations agree extremely well with the target dynamics. Thus, for the first time we have succeeded in quantitatively designing the transition kinetics of colloidal-like clusters, and can do so accurately over a wide range of rates. By optimizing over experimentally relevant parameters, these results give a direct prediction that can be tested in clusters of DNA-coated colloids.

\subsection{Simultaneous training of multiple transitions \label{sec:simultaneous_transitions}}
In the above results, transition kinetics are designed while simultaneously imposing specific structural constraints. While most self-assembly approaches focus on obtaining a specific structure, we design for both structure and kinetics simultaneously. Indeed, we have designed for {\it two} kinetic features, namely the forward energy barrier/rate and the backwards energy barrier/rate. Is it possible to take this further and simultaneously specify many kinetic features? Since particles are distinguishable, permutations represent distinct states and there are many thousands of possible transitions in our 7-particle clusters. What determines which subsets of transitions can be simultaneously and independently controlled?

To address this question, we develop an analogy to network rigidity. Each kinetic feature we would like to impose appears as a constraint in the loss function. If we want to impose $x$ kinetic features, then we attempt to minimize a loss function analogous to Eq.~\eqref{eq:barrierLoss} with $x$ terms of the form $(\Delta E - \Delta E^*)^2$.
\footnote{Or analogous to Eq.~\eqref{eq:rateLoss} with $x$ terms of the form $(k-k^*)^2$. For simplicity going forward, we will focus on imposing kinetic features via energy barriers.} 
This constraint satisfaction problem has a strong analogy to network rigidity~\cite{Jacobs:1995vi}, where physical springs constrain the relative positions of nodes. When attempting to place a spring, we can look to the eigenvalues of the Hessian matrix of the total energy to determine whether the new constraint can be satisfied. Each ``zero mode" (eigenvector of the Hessian with zero eigenvalue) represents a degree of freedom that can be adjusted slightly while maintaining all the constraints. The relevant question is whether the new constraint can be satisfied with only these free degrees of freedom. If this is the case, then imposing the constraint causes the system to lose a zero mode. However, if the additional constraint cannot be satisfied simultaneously with the other constraints, then the number of zero modes remains the same and the constraint is called ``redundant." Thus, if the number of non-zero modes equals the number of constraints, than all the constraints can be satisfied simultaneously. 

We employ this same approach in our system to determine if a set of constraints (desired kinetic features) can be simultaneously satisfied by a given set of variables (the 21 binding energies $B_{\alpha \beta}$). To proceed with a given set of desired kinetic features, we start with a random initial set of binding energies and write down a temporary loss function where the targets are equal to the current energy barriers. This is analogous to placing a spring between two nodes such that the spring's rest length is equal to the separation between the nodes. It also guarantees that the Hessian matrix $\frac{d^2L}{dB_{\alpha\beta} dB_{\alpha^\prime \beta^\prime}}$ is positive semi-definite and that the zero modes correspond to unconstrained degrees of freedom. We employ automatic differentiation to calculate the Hessian matrix, which we then diagonalize and count the number of zero and non-zero modes. If the number of non-zero modes equals the number of constraints, then we predict that the desired set of kinetic features can be simultaneously obtained. 

\begin{figure}[h!tpb]
	\centering
	\includegraphics[width=\linewidth]{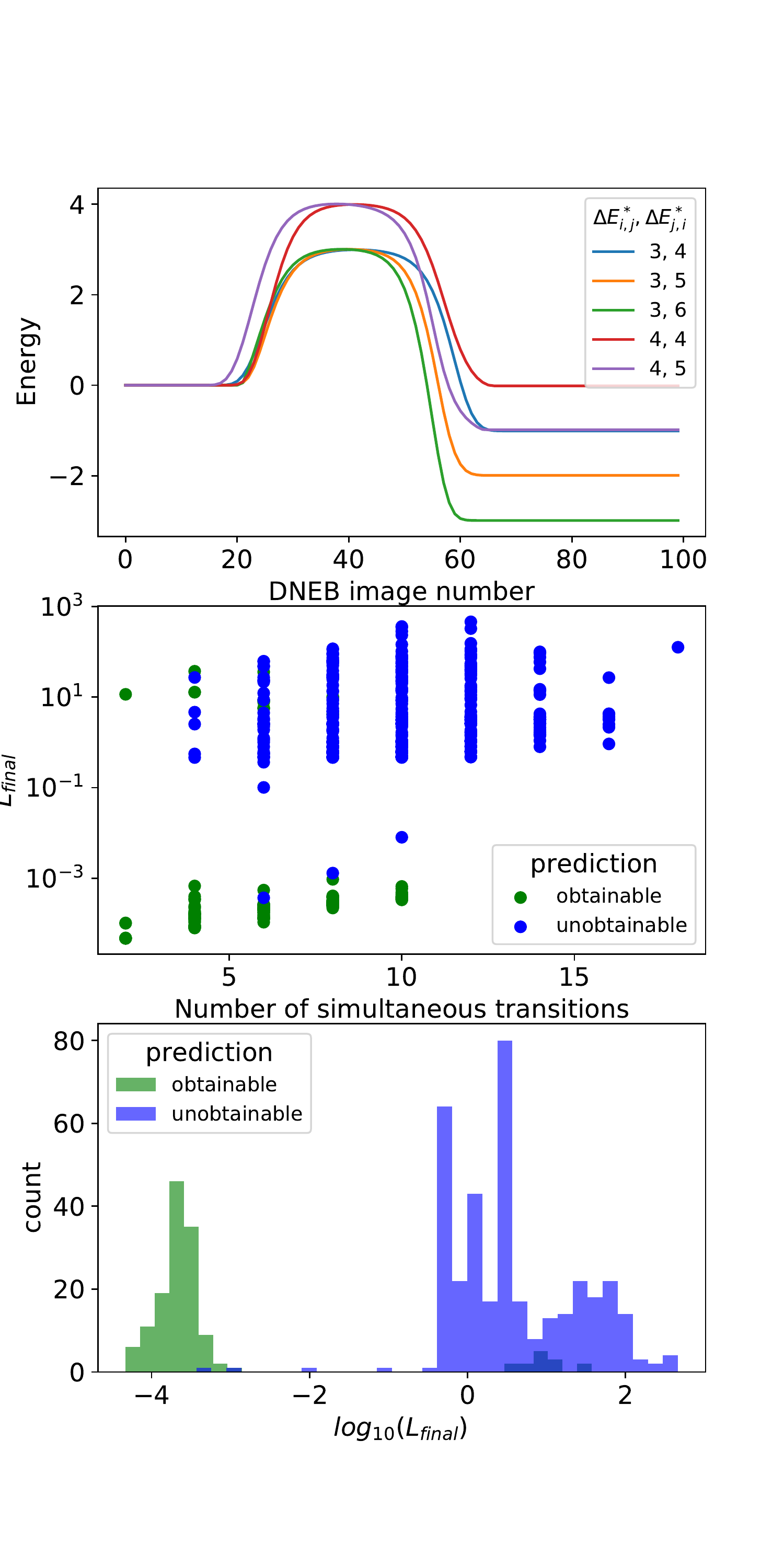}
	\caption{\label{fig:simultaneous_optimization}Energy (in $k_BT$) along the steepest descent paths connecting a single starting structure with five adjacent structures, obtained through the DNEB method (Appendix~\ref{sec:methods_transitions}). These results are after the simultaneous optimization of all 10 energy barriers (target energy barriers (in $k_BT$) are shown in the legend). The observed energy barriers are within $0.3\%$ of their targets. The inset shows the mean error for different number of transition rates being simultaneously optimized. 
	}
\end{figure}

To test this prediction, we have picked an initial structure and 9 adjacent structures, with the intention to simultaneously specify the energy barriers of various subsets of the 9 forward and 9 backward transitions. As a first example, we pick a set of 10 transitions (5 forward and 5 backward) where the Hessian has 10 non-zero modes, meaning our theory predicts that our optimization should be successful. This is confirmed in Fig.~\ref{fig:simultaneous_optimization}A, which shows the energy along the 5 transition pathways after simultaneously optimizing for all 10 barriers. The legend shows the target energy barriers, $\Delta E^*_{ij}$ and $\Delta E^*_{ji}$, for both the forward and backward transitions, respectively. Each observed barrier is within $0.3\%$ of the respective target and becomes more accurate with additional training using smaller training rates.

This result is generalized in Fig. \ref{fig:simultaneous_optimization}B and C, where we show the loss function $L_\mathrm{final}$ after optimization for all 511 subsets of the 9 transition pathways. Cases where we predict that the kinetic features can be simultaneously obtained (i.e. that $L_\mathrm{final}\rightarrow 0$) are shown in green while cases where we predict that the kinetic features cannot be simultaneously obtained are shown in blue. While our predictions are not perfect, we find that they are correct roughly $97\%$ of the time.

A key difference between our system and network rigidity is that we are trying to predict a highly non-local feature of the function space, namely whether or not a loss function can be minimized to zero. Conversely, rigidity in elastic networks is a local property of the energy landscape. We extrapolate from the local curvature to predict highly non-local behavior, without proper justification. The fact that this nevertheless works quite well indicates that there is indeed some persistent internal structure in the function space we are considering, despite its considerable complexity. We leave a more in-depth examination of this structure to future studies. 

\section{Discussion \label{sec:discussion}}
Rather than being limited to structural motifs, our results expand the design space of colloids with specific interactions to include complex kinetics. Understanding the connection between interactions and emergent dynamical properties is hugely important in biological physics, materials science, and non-equilibrium statistical mechanics. Moreover, achieving in synthetic materials the same level of complexity and functionality as found in biology requires detailed control over kinetic features.

We demonstrate quantitative control over kinetic features in two self-assembly systems. In the classic problem of assembling a honeycomb lattice using monotonic, spherically symmetric particles, we achieve nontrivial control over lattice assembly rates. Understanding and controlling crystallization kinetics is critical to the design of many real materials and is intimately related to the statistics of defects. Furthermore, since a honeycomb crystal is a diatomic lattice where neighboring particles have different orientational order, assembling a honeycomb lattice out of identical particles is a notoriously difficult problem. Our ability not only to assemble such a lattice but to control its assembly rate relative to a triangular lattice reveals an essential connection between assembly dynamics and particle interactions that we are able to exploit for kinetic design.

Secondly, by considering small clusters of colloidal-like particles with specific interactions, we discover that transition kinetics are far more designable than was previously thought. Since spherically symmetric particles provide a foundational understanding not only in self assembly but in fields throughout physics, our results suggest the same potential for designing transition dynamics in numerous other systems.

Moreover, the methodology we present is applicable beyond the study of colloidal kinetics. At its core, this method directly extracts the effect of experimental or model parameters on an emergent property, such as the kinetic rates that we focus on. It does so by exactly measuring the derivative of the emergent property with respect to the underlying parameters. We have shown that despite the stochasticity inherent to the thermal systems we consider, these derivatives are \textit{predictive}. In other words, the validity of the linear regime accessed by the gradient calculation is not overwhelmed by stochastic noise. Thus, we can cut through the highly complex dependence of emergent dynamical features on model parameters (such as interaction energies, size distributions, temperature and pressure schedules, or even particle shapes) and use this dependence to control behavior.

Thus, we have shown that gradient computations over statistical-physics calculations using automatic differentiation are possible, efficient, and sufficiently well behaved for optimization of kinetic features. Unlike for structural features, it is difficult to intuit the connection between interactions and kinetics. By revealing this connection, the differentiable statistical-physics models presented here provide a new framework for understanding emergent behavior. Indeed, our work merely scratches the surface of what can be achieved using the principles of kinetic design. This approach scales favorably with the number of input parameters, enabling the design of increasingly complex phenomena. We envision engineering materials with novel emergent properties, exploring the vast search space of biomimetic function, and manipulating entire phase diagrams. 

\section*{Acknowledgements}
We thank Agnese Curatolo, Megan Engel, Ofer Kimchi, Seong Ho Pahng, and Roy Frostig for helpful discussions. This material is based upon work supported by the National Science Foundation Graduate Research Fellowship under Grant No. DGE1745303. This research was funded by the National Science Foundation through DMS-1715477, MRSEC DMR-1420570, and ONR N00014-17-1-3029. MPB is an investigator of the Simons Foundation.

\section*{Author contributions} 
All authors designed the study and wrote the paper. C.P.G. and E.M.K. performed the numerical simulations and analysis, while S.S.S. and E.D.C. developed the underlying numerical package. C.P.G. and E.M.K contributed equally.

\section{Molecular Dynamics simulations \label{sec:Methods}}
We use Molecular Dynamics (MD) simulation in two ways. First, in Sec.~\ref{sec:honeycomb_crystallization} we directly differentiate through MD simulations in order to optimize the honeycomb and triangular lattice assembly rates. Secondly, in Sec.~\ref{sec:colloidal_clusters} we verify our optimized transition rates by running MD simulations and extracting transition rates from the simulations.

The MD simulations are performed using JAX MD~\cite{jaxmd2019,Schoenholz:2019tq}, a molecular dynamics engine that is compatible with JAX~\cite{jax}, a freely-available automatic differentiation library. The system consists of a two- or three-dimensional box containing $N$ particles at a constant temperature $T$. To simulate Brownian motion, the dynamics are given by the overdamped Langevin equation:
\begin{equation}\label{eq:overdampedLangevin}
\dot{r}_\alpha = \gamma^{-1} F_\alpha + \sqrt{2 k_BT \gamma^{-1}}f_\alpha(t)
\end{equation}
where $F_\alpha$ is the net force on particle $\alpha$, $\gamma=0.1$ is the friction coefficient, $k_B$ is the Boltzmann constant, and the elements of $f_\alpha(t)$ are uncorrelated Gaussian random variables with zero mean. 

In the case of optimizing lattice assembly rates, we simulate $N=100$ particles that interact via ~\eqref{eq:honpot}. The particles interact within a square two-dimensional simulation box with periodic boundary conditions and sides of length 11.4 and 9.31 for assembling the honeycomb and triangular lattices, respectively.  Simulations are performed at a temperature of $k_BT=0.1$ using a simulation step size of $5\times 10^{-5}$.

In the case of verifying transition rates, we simulate $N=7$ particles that interact via a Morse potential of the form
\begin{equation}\label{eq:morse}
V_{\alpha\beta}(r_{\alpha\beta}) = B_{\alpha\beta}\left( e^{ -2a(r_{\alpha\beta}-\sigma)} -2 e^{-a(r_{\alpha\beta}-\sigma)} \right)
\end{equation}

where $r_{\alpha\beta}$ is the separation between particles $\alpha$ and $\beta$, $\sigma=1$ is the particle diameter, $a=10$ determines the range of the attraction, and $B_{\alpha\beta}$ is the binding energy between the spheres. Simulations are performed at a temperature of $k_BT=1$ in a three-dimensional simulation box with free boundary conditions using a step size of $5\times 10^{-6}$.

\bibliography{mybibfile, papers} 

\section{Appendix}
\subsection{Optimizing lattice assembly rates (Sec.~\ref{sec:honeycomb_crystallization}) \label{sec:methods_crystallization}}
\subsubsection{Forward mode AD}

To optimize our system, we use the RMSProp stochastic optimizer ~\cite{rmsprop} with a learning rate of 0.1, a memory value of $\gamma_\mathrm{mem} = 0.9$, and a smoothing value of $\epsilon = 10^{-8}$. The optimizer acts on a gradient that is determined via an average of 100 independent simulations, each with random initial positions.

The loss function we use for the optimization consists of specifying a ``stencil", or a fragment of a perfect lattice centered around a central particle. The stencil was created with a particle diameter of 1. A honeycomb lattice stencil and a triangular lattice stencil are both specified. At the end of a simulation, we center the stencil on a particle $\alpha$ and define the overlap function
\begin{align}
O_\alpha(\theta) = \sum_{\beta, \beta^\prime} e^{- \frac{\left(r_\beta - r_{\beta^\prime}(\alpha,\theta)\right)^2}{2 \sigma^2}}
\end{align}

where $r_\beta$ is the position of particle $\beta$ and $r_{\beta^\prime}(\alpha,\theta)$ is the position of the $\beta^\prime$ particle in the stencil when the stencil is centered on particle $\alpha$ and rotated by an angle $\theta$. The maximum overlap of particle $\alpha$ is 
\begin{equation}
O_{\alpha,\mathrm{opt}} = \max_\theta O_\alpha(\theta)
\end{equation}
and we define total overlap of the system to be 
\begin{equation}\label{eq:omax}
O = \max_\alpha O_{\alpha,\mathrm{opt}.}
\end{equation}

\begin{figure}[h!tpb]
	\centering
	\includegraphics[width=\linewidth]{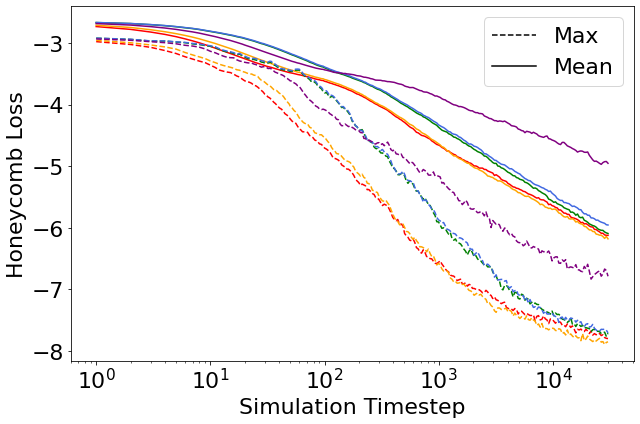}
	\caption{\label{fig:maxvmean} Assembly process for honeycomb lattices with 5 different sets of parameters, where a given set of parameters is given by a distinct color. The parameters are the same as those used to generate the data in Fig.~\ref{fig:kHkT_res}. The assembly process as measured by the maximum over $O_{\alpha, \text{opt}}$ (as shown in eq.~\ref{eq:omax})is given by dashed lines, whereas the assembly process as measured by the mean over $O_{\alpha, \text{opt}}$ is given by solid lines. We see that the two measurements show the same trend, and both serve as similar indicators of crystallization.}
\end{figure}

We use a large stencil, ensuring that a significant overlap is clearly indicative of a honeycomb-like region. Additionally, the particles all interact identically. These two features allow us to use the maximum overlap as an indicator of crystallization. The assembly process as measured by the maximum overlap is contrasted with the assembly process as measured by the mean overlap in Fig.~\ref{fig:maxvmean}.

We use a 13 particle stencil for the honeycomb overlap $O_{\text{hon}}$, and 7 particle stencil for the triangular overlap $O_{\text{tri}}$. These together comprise the two loss functions, weighted as follows:

\begin{equation}\label{eq:hon_loss}
L_H(t) = \frac{1}{N_H}(-O_{\text{hon}}(t) +  \xi_H O_{\text{tri}}(t)  + \zeta_H)
\end{equation}

\begin{equation}\label{eq:tri_loss}
L_T(t) = \frac{1}{N_T}(O_{\text{hon}}(t) - \xi_T O_{\text{tri}}(t) + \zeta_T)
\end{equation}

with $\xi_H = 1$ and $\xi_T = \frac{104}{7}$. We choose $\zeta_{H, T}$ and $N_{H, T}$ such that $L_H$ and $L_T$ range between 0 and 1, where $L_{H, T} = 0$ for a perfect lattice.

We perform two rounds of optimization. For input rates $k_H^* = 1/t^*_H$ and $k_T^* = 1/t^*_T$, the first round of optimization runs $t^*$ steps and  computes $L = (L_H(t^*_H) - 0.5)^2 + (L_T(t^*_T) - 0.5)^2$ for each of $t^*_H$ and $t^*_T$ under the appropriate density conditions. We then run the simulation for $t^*$ more steps and compute $L_H(2t^*_H)$ and $L_T(2t^*_T)$ at the end of both simulations. The optimization loss function is a sum of both losses, namely (1) how close the system is to half-assembled after $t^*$ steps and (2) how assembled the system is after $2t^*$ steps:

\begin{equation}
L_{\text{opt}} = \sum_{H, T}(L(t^*) - 0.5)^2 + L(2t^*)
\end{equation}

The optimal parameters are the parameters associated with the minimum loss value over 1,000 RMSProp optimization steps. We then perform a second optimization, starting with these parameters, in which the optimization loss is restricted to $L_{\text{opt}} = (L_H(t_H^*) - 0.5)^2 + (L_T(t_T^*) - 0.5)^2$.
In both rounds of optimization, we use forward mode automatic differentiation to calculate $\frac{d L_\text{opt}}{d a}$, where $a$ is the set of variable parameters in Eq.~\eqref{eq:honpot}.

To validate optimization results, we calculate the rates $k_\mathrm{H}$ and $k_\mathrm{T}$ by computing the losses $L_H(t)$ and $L_T(t)$ as a function of simulation time step.
We then find the earliest timestep, $t_0$, at which the loss is greater than half its optimal value, and the latest timestep, $t_1$, at which the loss is less than half its optimal value. Using data in the range $(t_0 - 3, t_1 + 3)$, we approximate a linear fit and use the fit to compute the time $t^{*}$ at which the loss function is exactly half its optimal value.

\subsubsection{Reverse mode AD}

Performing reverse mode AD is more time efficient, but uses more memory. To conserve memory, we present an indirect approach to optimizing lattice assembly rates in which we only differentiate over the final 300 time steps of the simulation. Consider breaking the simulation up into two components. We first simulate for $\tilde \tau$ time steps and find the configuration 
\begin{equation}
R_{\tilde \tau} = \mathcal{S}_\mathrm{MD} \left(a, R_0, \rho, \tilde \tau \right)
\end{equation}
at the end of the first simulation. We then begin from the configuration $R_{\tilde \tau}$ and simulate further for $\bar \tau = 300$ time steps, returning a final configuration 
\begin{equation}
R_{\bar \tau} = \mathcal{S}_\mathrm{MD} \left(a, R_{\tilde \tau}, \rho, \bar \tau \right). 
\end{equation}

Crucially, we differentiate over \emph{only the second simulation}. To optimize the loss at time $\tau=\tilde \tau + \bar \tau$, we calculate the derivative
\begin{equation}\label{eq:dLda}
\left. \frac{d \mathcal{L}\left( \mathcal{S}_\mathrm{MD}(a,R_{\tilde \tau} , \rho, \bar \tau) \right) } {d a}\right|_{R_{\tilde \tau}}. 
\end{equation}

Calculating this derivative at constant $R_{\tilde \tau}$ means that we only have to differentiate through $\bar \tau$ time steps. In practice, we find good results with $\bar \tau$ as small as $300$, which we hold constant while varying $\tilde \tau_\mathrm{H}$ and $\tilde \tau_\mathrm{T}$ to tune the two crystallization rates relative to each other. Using our results, we can interpolate and find a relationship between $\tilde \tau$ and the corresponding rate: $\tilde \tau$ is a knob we can tune to adjust the relative lattice assembly rates.

\subsection{Optimizing transition rates in colloidal clusters (Sec.~\ref{sec:colloidal_clusters})  \label{sec:methods_transitions}}
\subsubsection{The doubly nudged elastic band calculation}
To calculate the transition state between two known adjacent local minima in a high-dimensional energy landscape, we follow the procedure from Trygubenko and Wales~\cite{Trygubenko:2004ip}, which we briefly summarize here. We want to find the (monotonically increasing) steepest ascent path from the first minimum up to the transition state and the (monotonically decreasing) steepest descent path down to the second minimum.

Let $R_0$ and $R_{n_I+1}$ be the configurations of the two minima, and we will represent a path between the two as a series of $n_I$ configurations $\{R_1, R_2, ..., R_{n_I}\}$. As an initial guess, we always choose a simple interpolation between the two minima. Importantly, in order for this to be a reasonable guess, the two minima have to be rotated so that they are close to overlapping. The potential energy of the $i$-th individual configuration in the path is $U(R_i)$, which we refer to here as the ``true potential." Thus, the total true potential of the ensemble is
\begin{equation}
    V = \sum_{i=0}^{n_I+1} U(R_i).
\end{equation}

In addition, we connect each adjacent configuration with a high-dimensional spring, leading to the following ``elastic band" or ``spring" potential
\begin{equation}
\tilde V = \frac 12 k_\mathrm{spr} \sum_{i=1}^{n_I+1} \left|R_i - R_{i-1}\right|^2.
\end{equation}

In principle, one wishes to minimize $V_\mathrm{tot} \equiv V + \tilde V$ over all the $n_I$ intermediate configurations while keeping the two endpoints fixed at their respective minimum. However, interference between the true and spring potential can give rise to ``corner-cutting" and ``sliding-down problems." To address these problems, we employ a set of adjustments, or ``nudges," to the gradient of $V_\mathrm{tot}$, as follows.

{\it Nudging.} First, we decompose the gradient of each configuration into components that are parallel and perpendicular to the current path. Let $\guv \tau_i$ be the unit vector tangent to the path at configuration $i$, which is defined as follows. 
If configuration $i$ does not represent a local optimum, meaning exactly one of its neighbors $j$ has a higher energy, $U(R_i) < U(R_j)$, then its tangent vector is 
\begin{equation}
\guv \tau_i = \frac{(j-i)(R_j-R_i)}{\left|R_j - R_i\right|}.
\end{equation}
However, if either both or none of its neighbors are at higher energy, then we use
\begin{equation}
\guv \tau_i = \frac{R_{i+1} - R_{i-1}}{\left| R_{i+1} - R_{i-1} \right|}.
\end{equation}

The gradient of the true potential can then be decomposed into
\begin{align}
\v g_i = \v g_i^\parallel + \v g_i^\perp,
\end{align}
where
\begin{align}
\v{g}_i^\parallel &= \left(\nabla_i V \cdot \guv \tau_i\right) \guv \tau_i, \\
\v{g}_i^\perp &= \nabla_i V - \v g_i^\parallel.
\end{align}

Similarly, using tildes to denote quantities related to the spring potential,
\begin{align}
\v {\tilde g}_i = \v {\tilde g}_i^\parallel + \v {\tilde g}_i^\perp,
\end{align}
where
\begin{align}
\v{\tilde g}_i^\parallel &= \left(\nabla_i \tilde V \cdot \guv \tau_i\right) \guv \tau_i, \\
\v{\tilde g}_i^\perp &= \nabla_i \tilde V - \v{\tilde g}_i^\parallel. 
\end{align}

The nudged elastic band approach is to project out $\v g_i^\parallel$ and $\v {\tilde g}_i^\perp$ when minimizing $V_\mathrm{tot}$. This removes some but not all of the interference instabilities. The ``doubly nudged" approach is to only project out some of the $\v{\tilde g}_i^\perp$ term, so that 
\begin{equation}
\v g_i = \v g_i^\perp + \v{\tilde g}_i^\parallel + \v{\tilde g}_i^\perp - \left(\v{\tilde g}_i^\perp \cdot \v{\hat g}_i^\perp \right) \v{\hat g}_i^\perp.  
\end{equation}

We proceed by minimizing $V_\mathrm{tot}$ using this nudged gradient. We note that optimizing over such a connected ensemble is especially straightforward in JAX MD because automatic vectorization is natively built in. The result is a sequence of configurations that closely tracks the steepest descent path we are seeking. Furthermore, the image $R_t$ with the highest energy is an approximation of the true saddle point. 
Note that we do not refine $R_t$ using eigenvector following~\cite{Cerjan:1981jm,Wales:1989bl}, a practice that is necessary for many applications. While $R_t$ is therefore only an approximation, this seems to be adequate for our purposes.

\subsubsection{Optimization of transition kinetics}
As discussed in the main text, we optimize the transition kinetics by first calculating $R_t$ using the DNEB method, then calculating $\frac{d L}{d B_{\alpha\beta}}$ using backward mode automatic differentiation, where $L$ is the chosen loss function, and finally using this gradient to minimize $L$. 

The optimization is performed using the RMSProp algorithm as implemented in JAX~\cite{jax, rmsprop}. Note that after each step of the optimization, both the height $E_t$ and the position $R_t$ of the saddle point will change slightly. While it is possible to differentiate over the entire DNEB calculation, this is not necessary and we instead calculate $\frac{d L}{d B_{\alpha\beta}}$ at fixed $R_i$, $R_j$, and $R_t$. We furthermore find it unnecessary in practice to redo the DNEB calculation every optimization step. Instead, we take multiple optimization steps in between DNEB calculations, which increases the efficiency of the computation. We find that recalculating $R_t$ every 50 optimization steps   works well for this problem. Note that in Fig.~\ref{fig:dneb_image}B, the iteration number refers to the  number of times $R_t$ has been recalculated. 

We run a total of 18 such iterations. As before, we use a memory value of $\gamma_\mathrm{mem} = 0.9$, and a smoothing value of $\epsilon = 10^{-8}$. However, we use a variable learning rate of $0.064$, $0.016$, and $0.004$ for the first, second, and third set of 6 iterations, respectively. 

Finally, we note that due to the long-range tail in the potential (Eq.~\eqref{eq:morse}), the exact position of the two minima technically change slightly during optimization. Therefore, before calculating $R_t$ each iteration, we first recalculate the local minima, though in practice this does not make a significant difference. 

\subsubsection{Validation of transition rates using MD}
To validate the transition rates, we run 100 simulations (described above) for $3 \times 10^6$ time steps each. Every $10^4$ steps, we compare the positions of the particles to the $n_I$ images, calculated using the DNEB procedure, that compose the transition path between the two states. Specifically, we find the image $i$ that minimizes $\sum_{\alpha \beta} (r_{\alpha \beta} - r_{\alpha\beta, i})^2$, where $r_{\alpha \beta}$ is the distance between particles $\alpha$ and $\beta$ in the current state, and $r_{\alpha\beta, i}$ is the distance between the particles in image $i$. 

The grey signal in Fig.~\ref{fig:dneb_image}C shows $i_\mathrm{closest}$ as a function of time. Note that the images near 0 and 100 (roughly 0-20 and 80-100, see flat lines in Fig.~\ref{fig:dneb_image}B) are identical up to rotations, so the distinction between them is meaningless and we do not need to be concerned that the data in Fig.~\ref{fig:dneb_image}C does not reach $i_\mathrm{closest}=0$. Note also that this signal is quite noisy, due in part to fluctuations in directions that do not align with the transition path. Therefore, it is important to filter this signal to remove transients that do not correspond transitions between the two states. This is done with a second-order lowpass Butterworth filter (see black curve in Fig.~\ref{fig:dneb_image}C). 

This filtered signal is then matched to one of the two minima by comparing it to the image number corresponding to the transition state (purple dashed line in Fig.~\ref{fig:dneb_image}C), leading to a binarized signal (blue curve in Fig.~\ref{fig:dneb_image}C). We then calculate the average dwell time of each state, $\tau_i$ and $\tau_j$. The measured rates are then $k_{ij,\mathrm{MD}} = 1/\tau_{i}$ and $k_{ji,\mathrm{MD}} = 1/\tau_{j}$.

\begin{figure}[h!tpb]
	\centering
	\includegraphics[width=\linewidth]{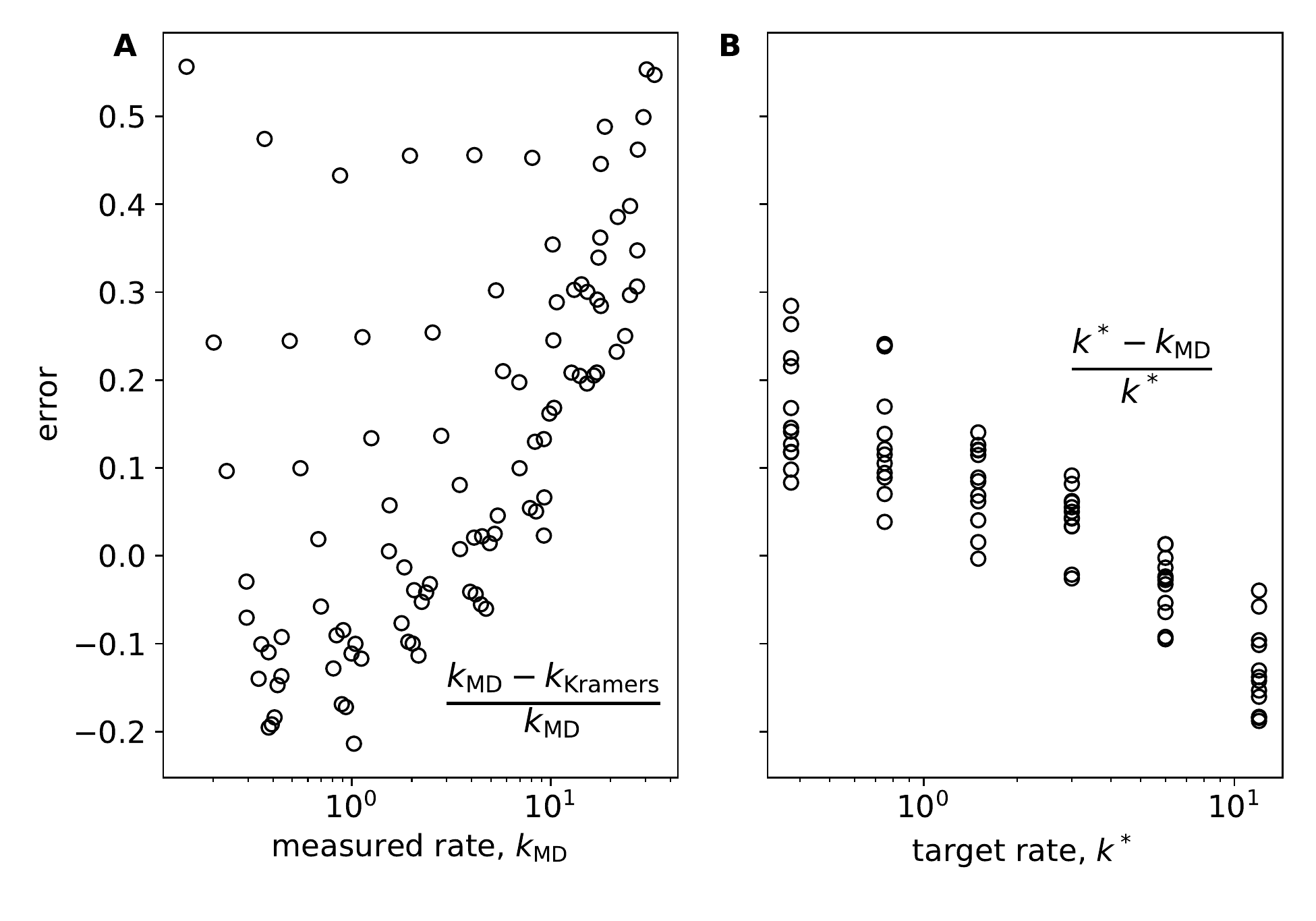}
	\caption{\label{fig:dneb_rate_errors} 
	Comparison of rates. A) Comparison of $k_\mathrm{Kramers}$ and $k_\mathrm{MD}$ when targeting desired energy barriers (see Fig.~\ref{fig:dneb_results}B). Perfect agreement is not expected because $k_\mathrm{Kramers}$ is an approximation that only considers the curvature at two points in the energy landscape, and the observed error of less than $50\%$ is very small compared to the two orders of magnitude of variation in the rates. B) Comparison of $k_\mathrm{MD}$ and the target rate $k^*$ when targeting desired transition rates (see Fig.~\ref{fig:dneb_results}D). Again, the observed error of mostly less than $20\%$ is expected and very small compared to the variation in the magnitude of the rates. The comparison of $k_\mathrm{Kramers}$ and $k^*$ is shown in Fig.~\ref{fig:dneb_results}C.} 
\end{figure}

Figure~\ref{fig:dneb_rate_errors} compares the rates extracted in this way to the rates obtained from the Kramer approximation (Eq.~\eqref{eq:Kramers}) and the target rate. 

\subsection{Using automatic differentiation in new systems}
In determining whether AD can be useful in studying a particular system of interest there are several considerations that must be taken into account. Given a function (that could involve an entire molecular dynamics simulation) that produces a scalar output, reverse-mode AD can compute its gradients using a single pass through the simulation. However, a consequence of this is that the entire simulation trajectory must be retained during the simulation. This induces a memory cost that grows both with the size of the simulation and the number of simulation steps and can quickly become unmanageable. There are several ways of ameliorating this cost. First, one can use gradient rematerialization to recompute short segments of the simulation during the backward pass. This typically reduces the memory cost to scale logarithmically in the duration of the simulation at the cost of a logarithmic increase in the required computational budget. Another option employed here is to use forward-mode AD which does not require additional storage during the simulation. However, here one pass through the entire simulation is required for each parameter and so the computational complexity grows quickly with the number of parameters in the function to be differentiated.  

\end{document}